\documentclass{aa}  
\usepackage{graphicx}
\usepackage{txfonts}
\usepackage{xcolor}

\begin{document}

   \title{Solar prominence diagnostics and their associated \\estimated errors from 1D NLTE \ion{Mg}{ii}~h\&k modelling}
   \author{A. W. Peat \inst{1,2},
          N. Labrosse \inst{2}, 
          K. Barczynski \inst{3,4}, \and
          B. Schmieder \inst{2,5,6}         
          } 

    \institute{University of Wroc\l{}aw, Centre of Scientific Excellence -- Solar and Stellar Activity, Kopernika 11, 51-622 Wroc\l{}aw, Poland\\\email{aaron.peat@uwr.edu.pl}
    \and
    SUPA School of Physics and Astronomy, The University of Glasgow, Glasgow, G12 8QQ, UK
    \and 
    ETH-Zurich, H\"onggerberg campus, HIT building, Z\"urich, Switzerland 
    \and 
    PMOD/WRC, Dorfstrasse 33, CH-7260 Davos Dorf, Switzerland 
    \and 
    LESIA, Observatoire de Paris, Universit\'e PSL , CNRS, Sorbonne Universit\'e, Universit\'e Paris-Diderot, 5 place Jules Janssen, 92190 Meudon, France
    \and 
    KU-Leuven, 3001 Leuven, Belgium}

   \date{}

 
  \abstract
   {}
   {We present further development of the rolling root mean square (rRMS) algorithm. These improvements {{consist of}} an increase in computational speed and an estimation of the uncertainty on the recovered diagnostics. {{This improvement is named the cross root mean square (xRMS) algorithm.}}}
   {We used the quantile method to recover the statistics of the line profiles in order to study the evolution of the prominence observed by IRIS on 1 October 2019. We then introduced the improvements to rRMS. These improvements greatly increased the computational speed, and this increase in speed allowed us to use a large model grid. Thus, we utilised a grid of 23~940 models to recover the thermodynamic diagnostics. We used the `good' (but not `best') fitting models to recover an estimate of the uncertainty on the recovered diagnostics.}
   {The maximum line-of-sight (LOS) velocities were found to be $70$~km~s$^{-1}$. {{The line widths were mostly 0.4~\AA{} with the asymmetries of most pixels around zero.}} The {central} temperature of the prominence was found to range from 10~kK to {20~kK}, with uncertainties of approximately {$\pm$5 to $\pm$15~kK}. The {central} pressure was around 0.2~dyn~cm$^{-2}$, with uncertainties of {$\pm$0.2 to $\pm$0.3~dyn~cm$^{-2}$}. The ionisation degree ranged from 1 to 1000, with uncertainties mostly in the range {$\pm$10 to $\pm$100}. The electron density was mostly {$10^{10}$~cm$^{-3}$}, with uncertainties of mostly {$\pm10^{9}$.}} 
   {The new xRMS algorithm finds an estimation of the errors of the recovered thermodynamic properties. To our knowledge, this is the first attempt at systematically determining the errors from forward modelling. The large range of errors found may hint at the degeneracies present when using a single ion and/or species from forward modelling. In the future, co-aligned observations of more than one ion and/or species should be used to attempt to constrain this problem.}

   \titlerunning{Solar prominence diagnostics and their associated estimated errors from \ion{Mg}{ii}~h\&k modelling}
   \authorrunning{A. W. Peat et al.}
   \maketitle
%

\section{Introduction}

Solar prominences are cool and dense chromosphere-like structures suspended in the hot and tenuous solar corona. {Prominence plasma typically have a temperature of 5000 to 8000~K, but there are reports of temperatures as low as 4300~K \citep{tandberg-hanssen_nature_1995}, and the edges of solar prominences have been seen to be of temperatures on the order of $10^4$~K \citep{hirayama_spectral_1971, hirayama_modern_1985}. {{This area of solar prominences}} has been interpreted as being the prominence-to-corona transition region (PCTR), where the plasma transitions from cool and dense prominence plasma to hot and tenuous coronal plasma. These temperatures are in stark contrast to the surrounding solar corona, which has temperatures on the order of $10^6$~K \citep{aschwanden_physics_2004}. }

The study of solar prominences can give us insight into the structure and thermodynamic properties of similar plasma and the wider solar atmosphere. It is therefore imperative that observations and modelling can be efficiently and effectively tied together in order to probe these enigmatic structures. {Many past and recent papers have attempted to discern the plasma parameters through the unification of observations and modelling. \cite{heinzel_SOHO_2001} performed qualitative comparisons with models and observations of the principal Lyman lines with the {{Solar Ultraviolet Measurements of Emitted Radiation instrument \citep[SUMER;][]{wilhelm_sumer_1995}}} aboard the Solar and Heliospheric Observatory \citep[SOHO; ][]{domingo_soho_1995}. Other studies have attempted to fit synthetic profiles with that of observations by reducing the line profiles to a handful of their statistics, such as their integrated intensities and full width half maximums \citep[e.g.][]{zhang_launch_2019, ruan_diagnostics_2019}. However, this method discards all the information about the shape of the profile. In response to this, \cite{peat_solar_2021} developed the rolling root mean square method (rRMS) in an attempt to match entire synthetic line profiles with observations.}

{One of the drawbacks inherent to all of these approaches is the lack of uncertainties.} In this paper, we present the cross root mean square method \citep[xRMS;][]{peat2023PhD}, an improvement of {rRMS}, to obtain diagnostics of {the prominence plasma and additionally to recover} some estimate of the errors on {the} acquired diagnostics. The first section of this paper concerns the observations of the prominence and what information may be gleaned from them, focusing on the observed dynamics from imaging telescopes. {This provides context for the following sections.} The second section focuses on the statistics that can be drawn from the line profiles with the use of the quantile method. These statistics are interpreted to understand the spatial and temporal evolution of the prominence plasma.
The third section concerns a summary and demonstration of the new xRMS procedure and how it is able to recover estimated errors. A discussion of what the recovered estimated errors imply about the use of a single ion and/or species from 1D prominence modelling to recover the thermodynamic properties of an observation.
Finally, we offer some brief conclusions on the findings and consequences of the paper.

\section{Observations}
\begin{figure}
    \centering
    \includegraphics[width=\linewidth]{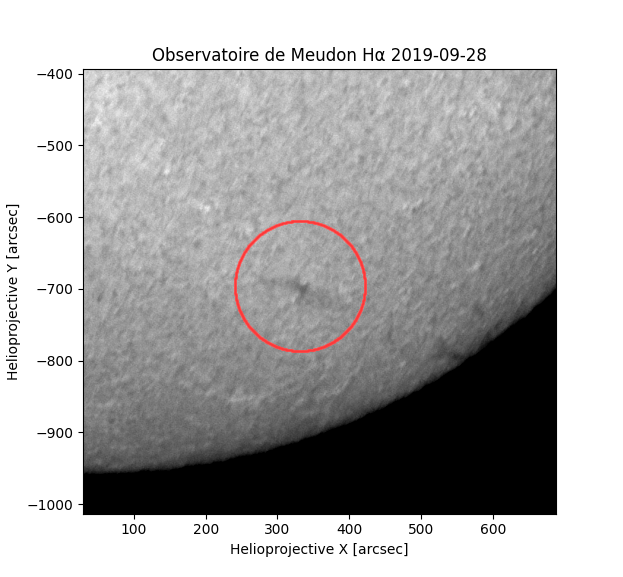}
    \caption{Prominence as it appeared as a filament in H$\alpha$ on 28 September 2019. The filament is highlighted with a red circle.}
    \label{fig:meudon}
\end{figure}

\begin{figure}
    \centering
    \includegraphics[width=\linewidth]{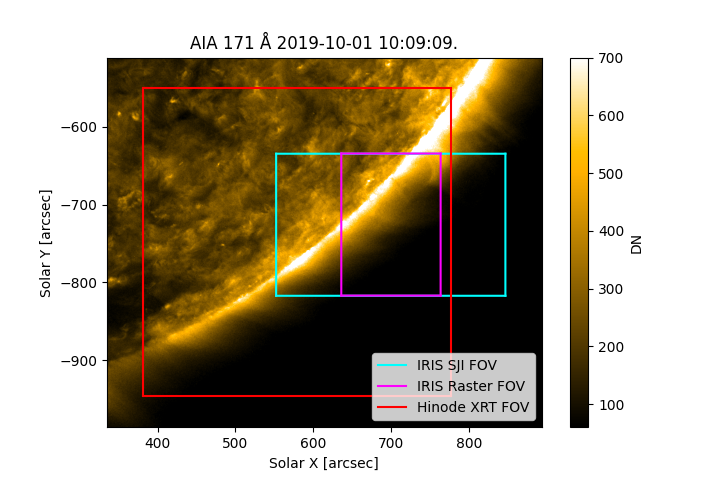}
    \caption{Configuration of this coordinated observation. The pointing of THEMIS is not trivial to determine, so it is omitted from this plot.}
    \label{fig:fovs}
\end{figure}

A filament appeared in H$\alpha$ observations of the Meudon Spectroheliograph\footnote{http://bass2000.obspm.fr/} on 28 September 2019. This later manifested as a prominence of the southwestern solar limb on 1 October 2019 (see Fig. \ref{fig:meudon}).
This prominence was observed with the Interface Region Imaging Spectrograph \citep[IRIS;][]{depontieu_interface_2014} as part of a coordinated observation with Hinode \citep{kosugi_hinode_2007} and the T\'{e}lescope Heliographique pour l'Etude du Magn\'{e}tisme et des Instabilit\'{e}s Solaries \citep[THEMIS;][]{mein_themis_1985}. {{For the THEMIS observation, see \citet{barczynski_two_2023}}}. The IRIS observations are composed of a very large, coarse 64-step raster. The observation began at 10:09~UTC and ended at 14:49~UTC with a field of view (FOV) of 127.71\arcsec$\times$182.32\arcsec centred on helioprojective coordinates 700.00\arcsec, -725.80\arcsec. The observation was composed of only the \ion{Mg}{ii} (2790.60--2809.90~\AA) passband and its complimentary slit-jaw imager passband (2796$\pm$2\AA). The X-Ray telescope \citep[XRT; ][]{golub_x-ray_2007} on board Hinode had an FOV of 394.98\arcsec$\times$394.98\arcsec, centred on helioprojective coordinates 579.42\arcsec, -748.54\arcsec. Images from the Atmospheric Imaging Assembly \citep[AIA; ][]{lemen_atmospheric_2012} on board the Solar Dynamics Observatory \citep[SDO;][]{pesnell_solar_2012} are also available. This configuration can be seen in {{Figure}} \ref{fig:fovs}.

\subsection{XRT}
\begin{figure}
    \centering
    \includegraphics[width=\linewidth]{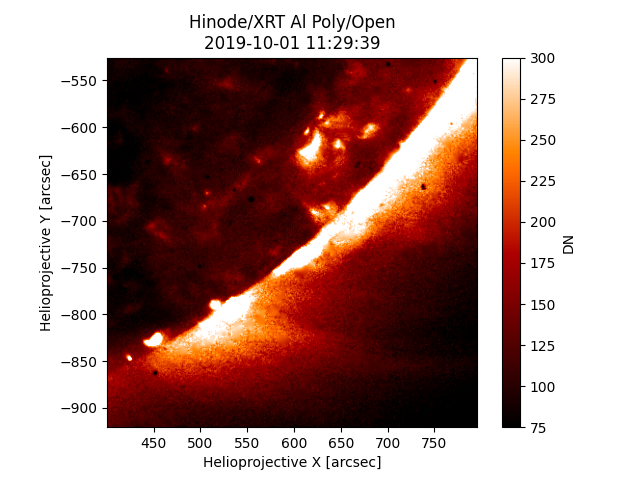}
    \caption{Hinode/XRT Al Poly/Open filter observations. The coronal cavity is clearly visible around 700\arcsec, -800\arcsec.}
    \label{fig:xrt}
\end{figure}

The XRT data was downloaded from the Hinode Science Data Centre Europe, and the dust spots were removed using \texttt{XRT\_SPOTCOR.pro} from SolarSoft \citep[SSW;][]{freeland_data_1998}. The coronal cavity in which the prominence resides is clearly seen in {{Figure}} \ref{fig:xrt}.  However, over the course of the observation, the XRT images display no significant changes in morphology or brightness.

\subsection{IRIS}
\begin{figure*}
    \centering
    \resizebox{\hsize}{!}
    {\includegraphics[width=0.33\linewidth]{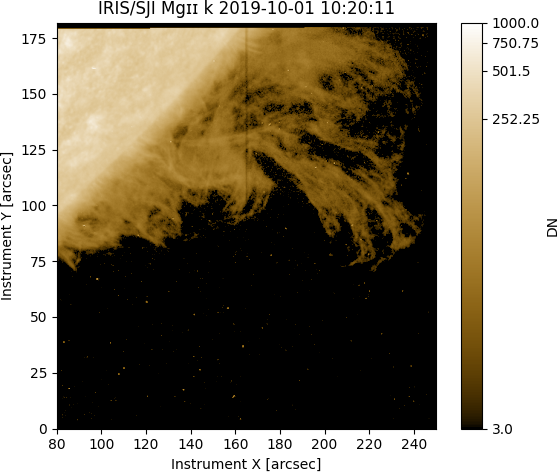}
    \includegraphics[width=0.33\linewidth]{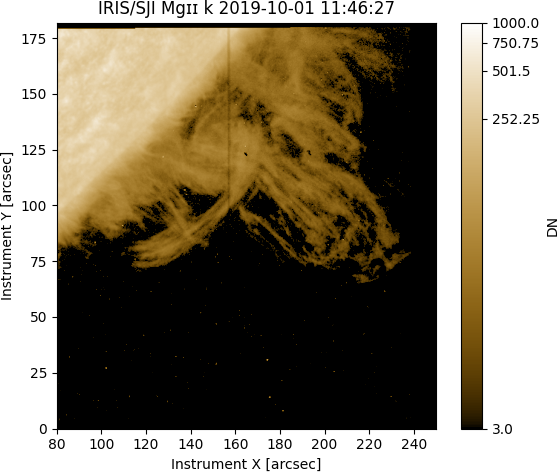}
    \includegraphics[width=0.33\linewidth]{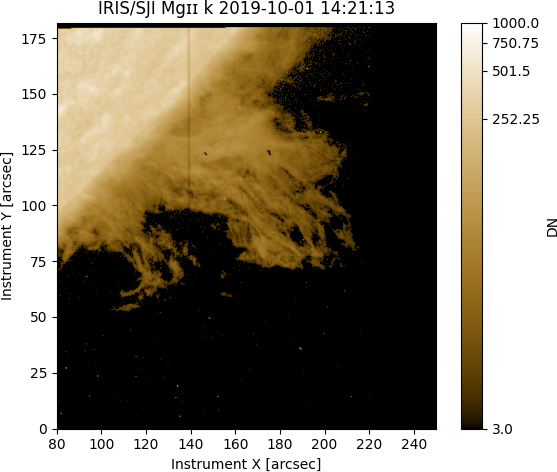}
    }
    \resizebox{\hsize}{!}
    {\includegraphics[width=0.33\linewidth]{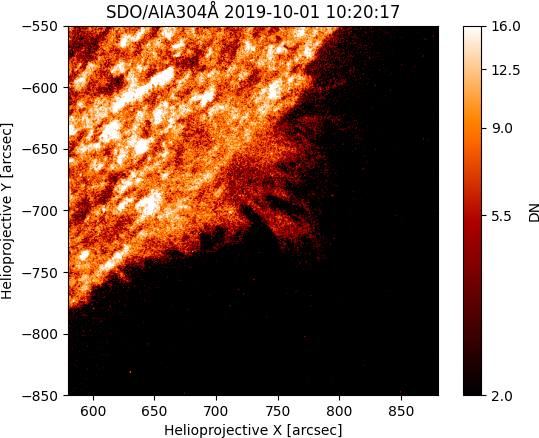}
    \includegraphics[width=0.33\linewidth]{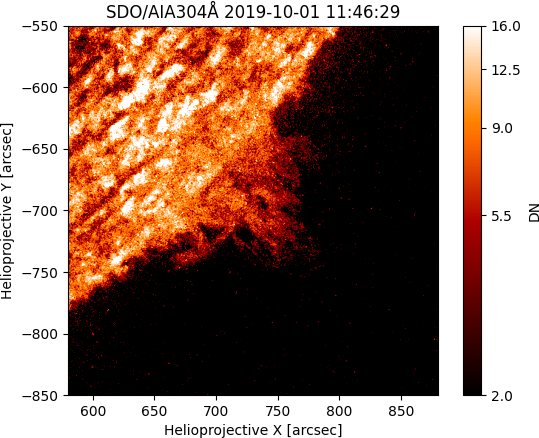}
    \includegraphics[width=0.33\linewidth]{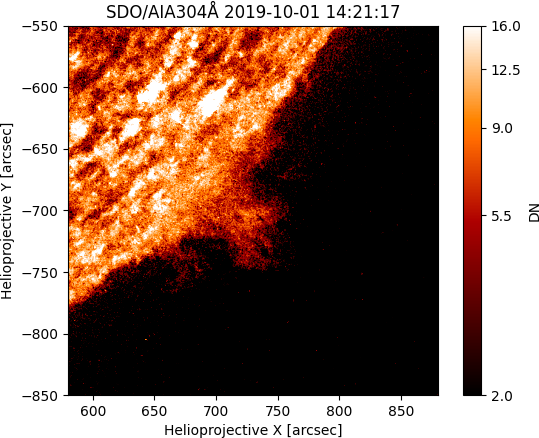}
    }
    \resizebox{\hsize}{!}
    {\includegraphics[width=0.33\linewidth]{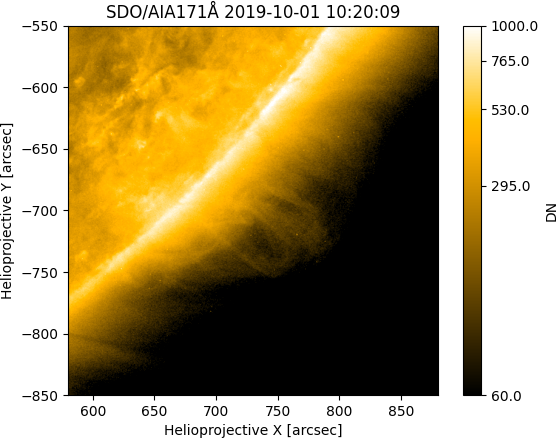}
    \includegraphics[width=0.33\linewidth]{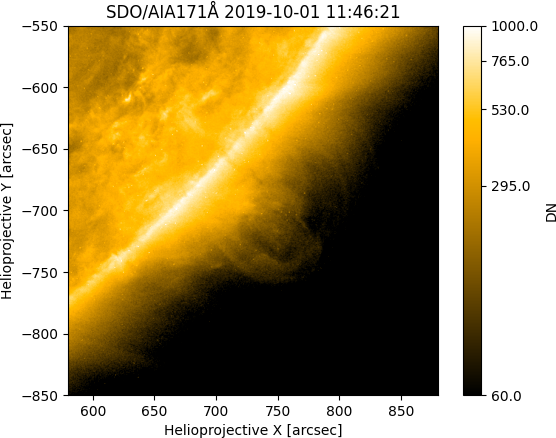}
    \includegraphics[width=0.33\linewidth]{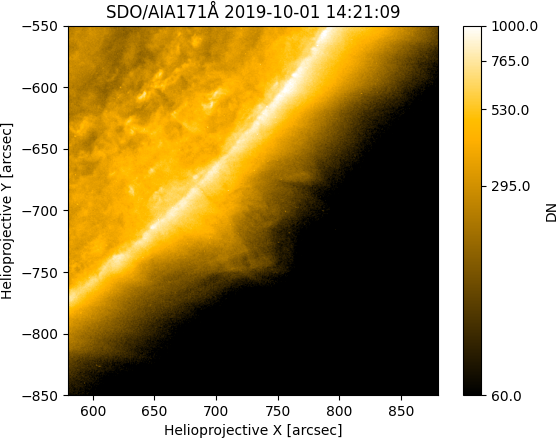}
    }
    \caption{Beginning, middle, and end (from left to right) of the IRIS prominence observation. The top, middle, and bottom rows are \ion{Mg}{ii}~k, 304~\AA\, and 171~\AA, respectively. The prominence appears far more extended in \ion{Mg}{ii}~k than in 304~\AA{}, but the IRIS/SJI FOV does not cover the entire structure as AIA does. Notably, 171~\AA{} allows us to see the high temperature PCTR that shrouds the structure.}
    \label{171304}
    
\end{figure*}
The IRIS raster files were retrieved from the Lockheed Martin Solar and Astrophysics Laboratory (LMSAL) as level 2 FITS files. These files were then radiometrically calibrated and deconvolved using the \texttt{irispreppy} package.\footnote{Available at https://github.com/OfAaron3/irispreppy or via pip.} Coronal and disc pixels were removed through the use of the line width filter described in \cite{peat_solar_2021}. This process works by calculating the line width of every pixel via the quantile method (see Sect. \ref{spec}) and plotting these values as a histogram. This produces a double-peaked distribution. Pixels in the corona (i.e. with no \ion{Mg}{ii} emission) are measured to have a large line width by the quantile method, so {{it is assumed}} that pixels with a large line width are coronal. Any pixel {{with}} a line width greater than the turning point of the line width distribution {{becomes a candidate for removal}}. To combat eliminating pixels that truly do contain data with a large line width, these candidate pixels for removal are {{then}} subjected to a simple intensity filter, and if they are above some threshold intensity, they are not eliminated. A third step \citep[described in][]{peat2023PhD} aims to eliminate single pixels that slip through the filter. In this step, the filter applies the `death by underpopulation' rule from Conway's Game of Life \citep{gardner_mathematical_1970}; any unremoved pixel with two or less unremoved {{2-connected}} pixels is itself removed.

\subsection{AIA}
The AIA files were retrieved from the Joint Science Operations Center (JSOC) as level 1 FITS files. These were then prepared to level 1.5 with the use of the \texttt{aiapy} package \citep{aiapy}. Figure \ref{171304} shows three different snapshots of the prominence evolution. This prominence does not appear as dynamic in 171~\AA{} and 304~\AA{} as it does in 2796~\AA{}, but the prominence is still clearly visible. In 304~\AA{}, we observed a similar overall structure to that seen in 2796~\AA{}, but the fine structure is not well resolved. The flow of material from the main body in the prominence, however, is visible. In 171~\AA{}, we would expect to be able to see the cool barb that anchors the prominence to the solar surface \citep[e.g.][]{parenti_nature_2012}, and either it has rotated out of or  it is yet to rotate into view. However, the `shroud' of the PCTR is easily seen here.

\section{IRIS spectra}

\begin{figure}
    \centering
    \includegraphics[width=.49\linewidth]{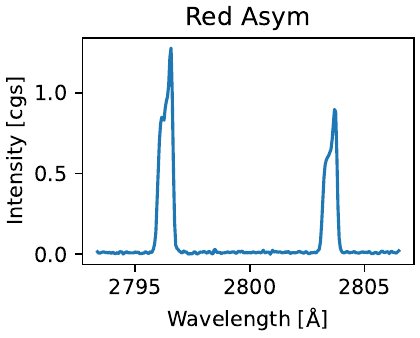}
    \includegraphics[width=.49\linewidth]{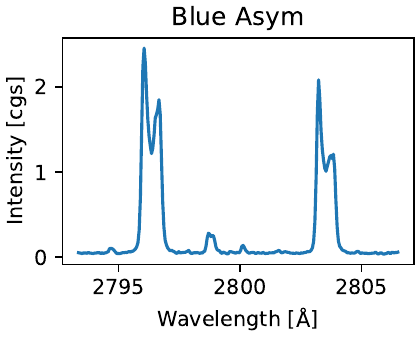}
    \caption{Examples of red (left) and blue (right) asymmetry. The intensities here are measured in  $10^4$~erg~s$^{-1}$~cm$^{-2}$~\AA{}$^{-1}$~sr$^{-1}$.}
    \label{fig:asymex}
\end{figure}

\begin{figure*}
    \centering
    \resizebox{0.8\hsize}{!}
    {\includegraphics{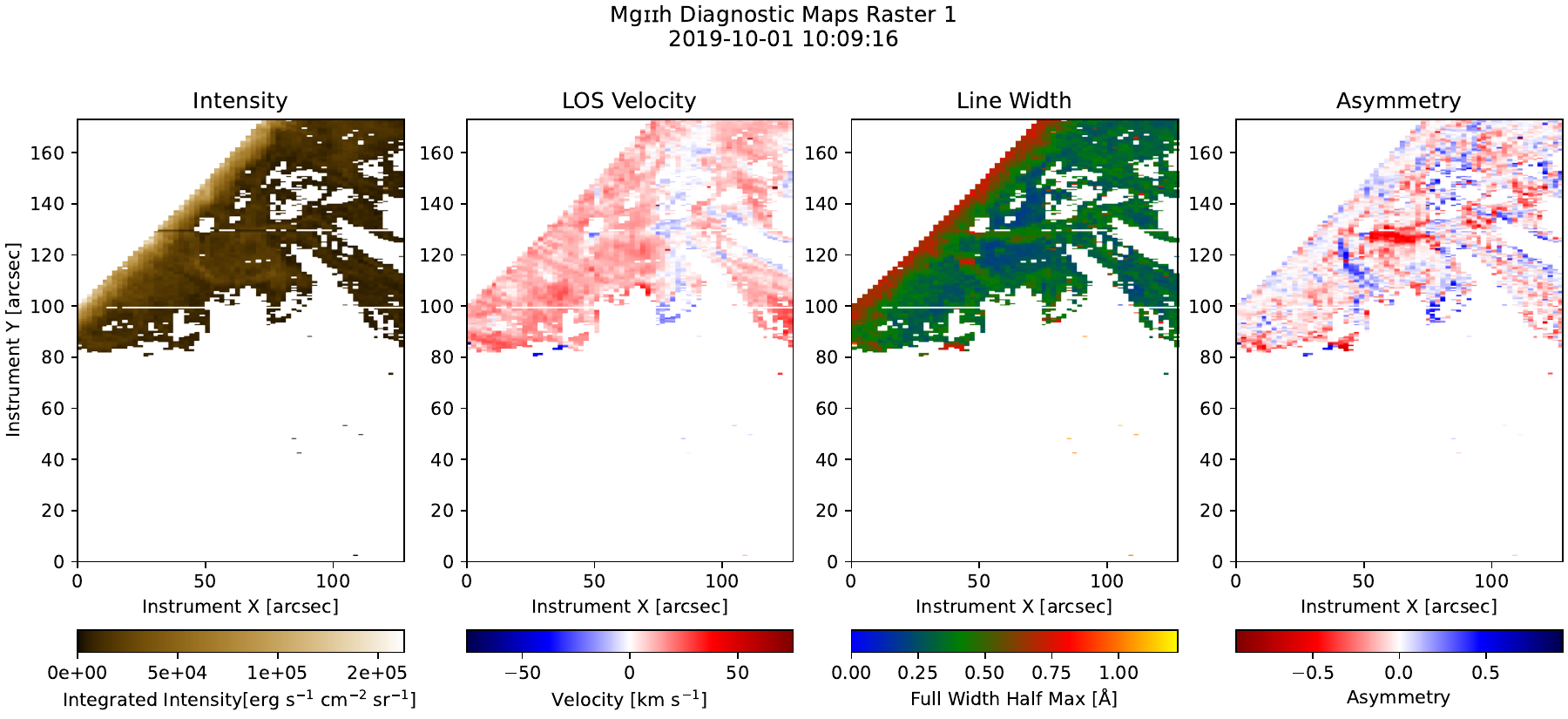}}
    \resizebox{0.8\hsize}{!}
    {\includegraphics{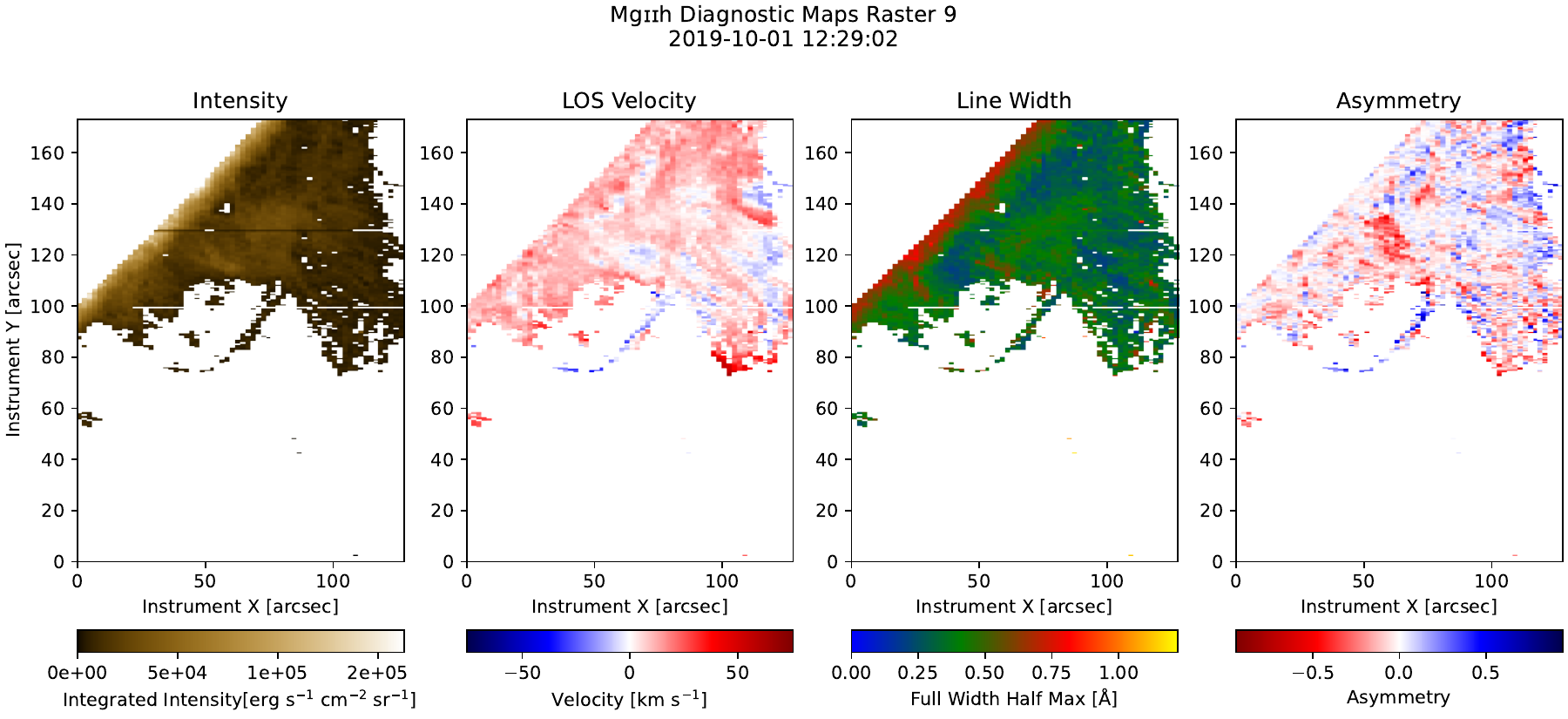}}
    \resizebox{0.8\hsize}{!}
    {\includegraphics{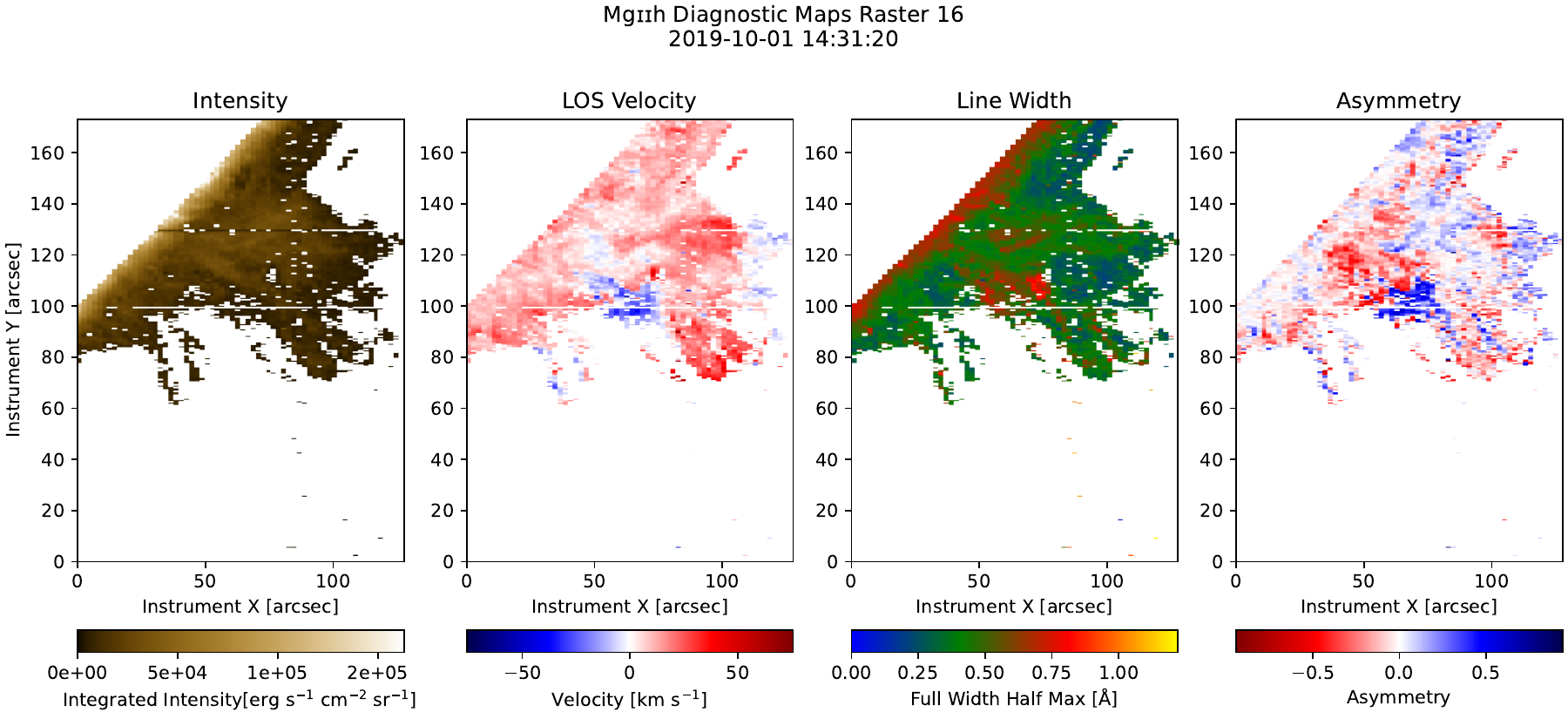}}
    \caption{Maps of the \ion{Mg}{ii}~h line profile statistics from the start, middle, and end of the observation. A similar plot for \ion{Mg}{ii}~k may be seen in Fig. \ref{kmaps}.}
    \label{maps}
\end{figure*}

\begin{figure}
    \centering
    \includegraphics[width=\linewidth]{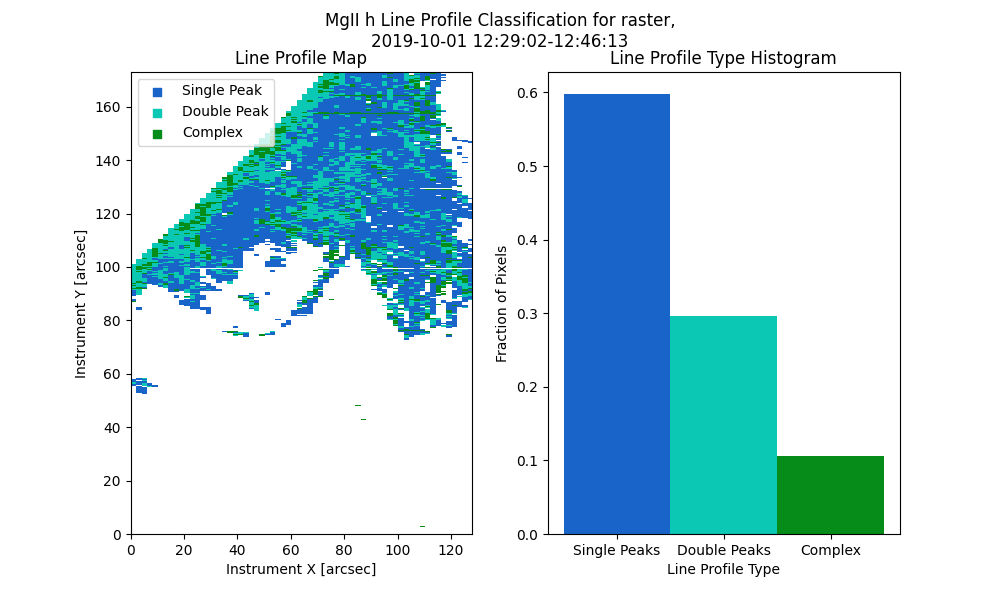}
    \includegraphics[width=\linewidth]{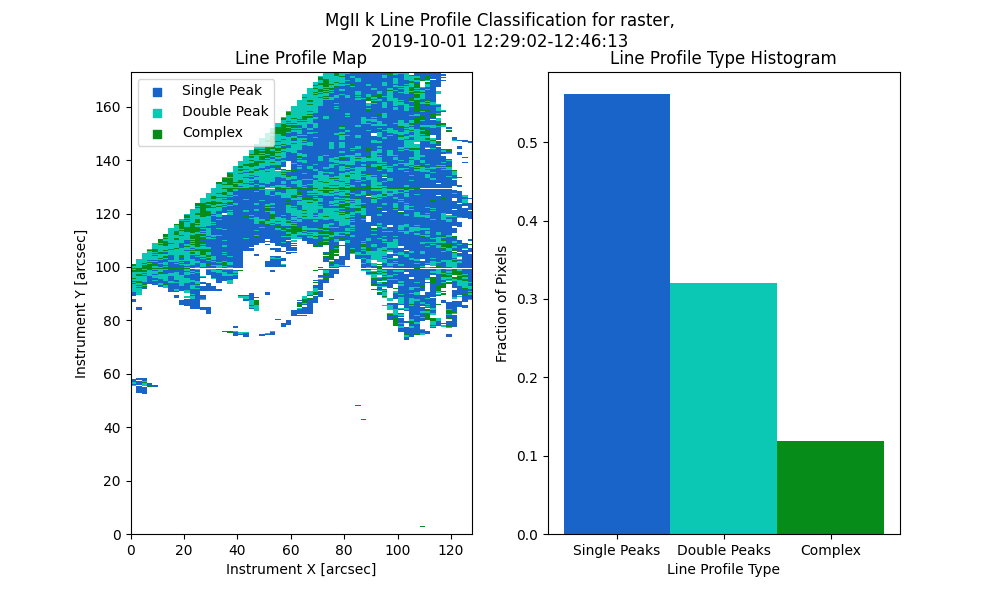}
    \caption{Line profile distribution of raster {9} of the prominence of 1 October 2019.}
    \label{fig:20191001lp}
\end{figure}
\label{spec}
To recover some statistics and diagnostics of the line profile, we employed the quantile method \citep{kerr_iris_2015,ruan_dynamic_2018}. A 3\AA{} window centred on the rest wavelengths of the \ion{Mg}{ii}~h\&k lines, respectively, was used to isolate each line. From this, a cumulative distribution function (CDF) was calculated for each pixel and normalised such that the values in the function range from zero to one. The wavelength at which the 50\% level of the CDF is found ($\lambda_{50}$) was defined as the line core. If the line is Gaussian in shape, the difference between $\lambda_{88}$ and $\lambda_{12}$ is the full width half maximum. However, it is understood that the lines are not all of a Gaussian shape, so we instead refer to this as the line width. Additionally, we calculated the asymmetry by the following relationship:
\begin{equation}
    \text{Asymmetry}=\frac{\left(\lambda_{88}-\lambda_{50}\right)-\left(\lambda_{50}-\lambda_{12}\right)}{\lambda_{88}-\lambda_{12}}=\frac{\lambda_{88}-2\lambda_{50}+\lambda_{12}}{\lambda_{88}-\lambda_{12}},
\end{equation}
where a positive number indicates there is more emission in the blue and a negative number indicates there is more emission in the red {(see Fig. \ref{fig:asymex})}. Figure \ref{maps} shows these statistics for the beginning, middle, and end of the observation. At the beginning, a structure with a large red asymmetry between 50\arcsec\ and 75\arcsec can be seen. It appears to lie between the very low values (<0.02\AA) in the line width map. It is only later, at the end of the observation, that these regions of large asymmetry begin to display a large line width. When both of these values are high, it implies the {{existence of fine structure}} in that pixel. The quantile method cannot discern the difference between one or two line profiles in one pixel, so it measures them as if they were one {\citep{peat2023PhD}}. This is due to the nature of using a CDF over some wavelength range centred on the rest wavelength of the line. In the middle of the observation, we observed a flow of material with a strong blueshift moving towards the observer at an excess of 60~km~s$^{-1}$. Another such flow appeared to begin near the end of the observation, with some material at 75\arcsec, 100\arcsec{} and {{also appeared}} highly blueshifted. However, this area also exhibits a large line width and blue asymmetry. Therefore, this blueshifted velocity could be due to the `asymmetry bias' outlined in \cite{peat_solar_2021}, which could be caused by more than one line profile being visible in that area.

\section{Forward modelling}

We used the 1D non-local thermodynamic equilibrium (NLTE) radiative transfer code PROM \citep{gouttebroze_hydrogen_1993, heinzel_theoretical_1994}. In the two introductory PROM papers, the prominence was represented by a {1D slab-like} isothermal and isobaric structure suspended in the solar corona with a focus on the six principal hydrogen transitions. \cite{labrosse_nonlte_2004} introduced a prominence-to-corona transition region (PCTR) in order to study the \ion{He}{i} triplet lines, which require a PCTR to form. \cite{levens_modelling_2019} introduced a \ion{Mg}{ii} ion to produce the \ion{Mg}{ii}~h\&k and \ion{Mg}{ii} triplet lines, with incident radiation taken from Sun-centre quiet Sun observations by IRIS. {The incident radiation is the angle-averaged irradiation on both sides of the prominence. The incident radiation used in this study is that used in \cite{levens_modelling_2019}, as the grid of models used here is an expansion built on the grid of 1007 models presented in the aforementioned study.}

With PROM, we generated a grid of 23~940 models, the parameters of which can be seen in Table \ref{params}. These models are a mix of isobaric and isothermal atmospheres and atmospheres containing a PCTR. The terms $T_{\text{cen}}$ and $p_{\text{cen}}$ are the central temperature and pressure, respectively; $T_{\text{tr}}$ and $p_{\text{tr}}$ are the temperature and pressure at the edge of the PCTR, respectively; slab width is the width of the slab; $M$ is the column mass; $H$ is the height above the solar surface; $v_T$ is the microturbulent velocity; $v_{\text{rad}}$ is the velocity of the slab parallel to the normal of the solar surface; and $\gamma$ is a dimensionless number that dictates the extent of the PCTR. For isothermal and isobaric models, there is only one temperature and pressure, $T=T_{\text{cen}}$ and $p=p_{\text{cen}}$. For non-zero values of $\gamma$, the higher the value of $\gamma$, the steeper the temperature gradient in the PCTR. The PCTR is formulated as a function of column mass \citep{kippenhahn_eine_1957, anzer_energy_1999},
\begin{equation} 
    T(m)=T_{\text{cen}}+(T_{\text{tr}}-T_{\text{cen}})\left(1-4\frac{m}{M}\left(1-\frac{m}{M}\right)\right)^\gamma,
    \label{tstrat}
\end{equation}
\begin{equation}
    p(m)=4p_c\frac{m}{M}\left(1-\frac{m}{M}\right)+p_{\text{tr}},
    \label{pstrat}
\end{equation}
for $\gamma\geq2$ and where $p_c=p_{\text{cen}}-p_{\text{tr}}$. A $\gamma$ value of zero indicates the model is isothermal and isobaric; this is merely a placeholder value and is not used in Eqs. \ref{tstrat} and \ref{pstrat}. 

Before these model profiles could be matched, they had to first be interpolated down to match the resolution of the IRIS spectrograph. There were two candidates for this interpolation: standard linear and fourth-order weighted essentially non-oscillatory interpolation \citep[WENO4;][]{janett_novel_2019}. Both of these schemes were tested, and they produced similar or identical results. Ultimately, linear was chosen for its expedience. More correctly, one would instead resample the data, but an interpolation is a good approximation.
\begin{table}
\caption{Model parameters. We note that not all of these {{models parameters}} are uniquely combined.}
\begin{tabular}{lcc} \hline\hline
Parameter        & Unit           & Value                                                                                                                                                      \\
\hline
$T_{\text{cen}}$ & kK              & \begin{tabular}[c]{@{}c@{}}6, 8, 10, 12, 15\\ 20, 25, 35, 40\end{tabular} \\
$T_{\text{tr}}$  & kK              & 100                                                                                                                                              \\
$p_{\text{cen}}$ & dyne cm$^{-2}$ & \begin{tabular}[c]{@{}c@{}}0.01, 0.02, 0.05\\ 0.1, 0.2, 0.5, 1 \end{tabular}\\
                                                                                                                       
$p_{\text{tr}}$  & dyne cm$^{-2}$ & 0.01                                                                                                                                                       \\
Slab Width       & km             & 45 -- 124~100                                                                                                                                         \\
M                & g cm$^{-2}$    & 3.7$\times10^{-8}$ -- 5.1$\times10^{-4}$                                                                                                                   \\
H                & Mm             & 10, 30, 50                                                                                                                                        \\
$v_{\text{T}}$   & km s$^{-1}$    & 5, 8, 13                                                                                                                                                   \\
$v_{\text{rad}}$ & km s$^{-1}$    & \begin{tabular}[c]{@{}c@{}}0, 2, 4, 6, 8, 10, 20\\40, 60, 80, 100, 150, 200 \end{tabular}\\

$\gamma$         &                & 0, 2, 4, 5, 10        \\ \hline \hline                                                                                                                    
\end{tabular}\\
{Notes:} $T_{\text{cen}}$ and $p_{\text{cen}}$ are the central temperature and pressure, respectively; $T_{\text{tr}}$ and $p_{\text{tr}}$ are the temperature and pressure at the edge of the PCTR, respectively; slab width is the width of the slab; $M$ is the column mass; $H$ is the height above the solar surface; $v_T$ is the microturbulent velocity; $v_{\text{rad}}$ is the velocity of the slab parallel to the normal of the solar surface; and $\gamma$ is a dimensionless number that dictates the extent of the PCTR. A $\gamma$ value of zero is mathematically meaningless, and it is simply used as a placeholder to identify isothermal and isobaric models.
\label{params}
\end{table}

\section{xRMS}
\begin{figure}
    \centering
    \includegraphics[width=\linewidth]{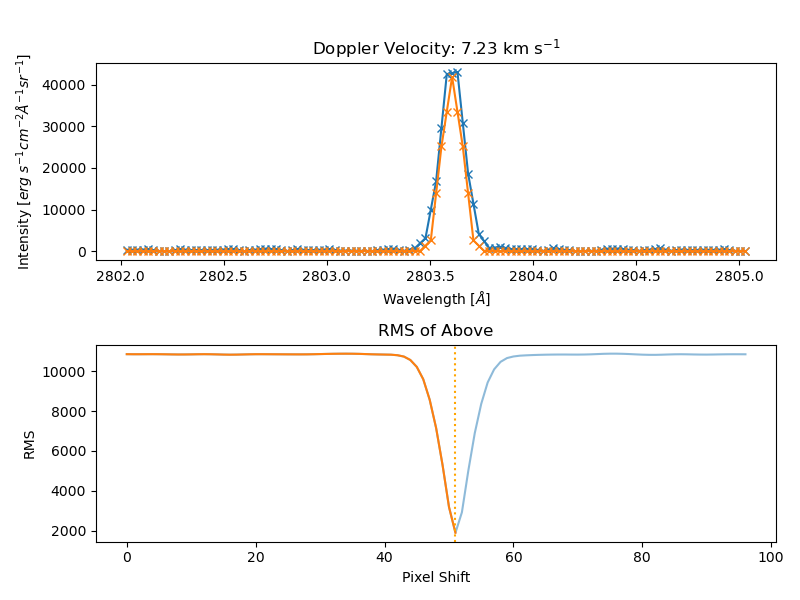}
    \caption{How rRMS worked. The orange model profile is rolled over the blue observed profile while measuring the RMS and pixel shift (and consequently Doppler velocity) at every wavelength position. {This process has been replaced by a cross-correlation in xRMS.} This figure is adapted from \cite{peat2023PhD}.}
    \label{fig:rrms}
\end{figure}
{The xRMS method} is an upgraded version of {the rRMS method \citep{peat_solar_2021}}. The rRMS method works by measuring the RMS between the models and the whole raster at once by vectorising the problem. A simple one-pixel example of how it works can be seen in Figure \ref{fig:rrms}. The RMS is measured at every position along the wavelength window, shifting it one pixel at a time. The pixel shift that produces the lowest RMS is selected as the ideal pixel shift for that model. The RMS and associated pixel shift are saved, and then the next model is run. {This process is repeated for every model, and} the model that produces the lowest RMS from this process is selected as the best model. {However, the best-fitting model is not necessarily a good fitting model. For a model to be classified as a `satisfactory match', its RMS must be below a threshold of 15~000 \citep{peat_solar_2021}.}

However, a cross-correlation is naturally produced when attempting to minimise (or maximise) a mean square \citep{elliott_handbook_1987} and, therefore, also a root mean square. Additionally, \cite{brown_doppler_2016} demonstrated the use of a cross-correlation to measure the line core shift of the Lyman lines in solar flares. Since measuring the pixel shift {in wavelength space} is essentially measuring the line core shift, {{we could instead use a cross-correlation}}. Using this approach, we were able to know the ideal pixel shift ahead of time and could skip straight to measuring the RMS at that position {only}. This improved run time by a factor of ten. However, the models still had to be rolled to their {ideal} positions, which is a computationally expensive procedure. This was replaced by instead padding the model with zeros and taking slices of the padded array to produce what would be the result of the roll. This resulted in a factor of two increase in speed, giving us an overall improvement of a factor of 20. In \cite{peat_solar_2021}, who {{introduced}} rRMS, one raster (of the size 32$\times$520) for 1007 models had a run-time of approximately 90 minutes on a dual socket motherboard with two Intel Xeon E5-2697 v4s. With {{the}} above-mentioned improvements, {xRMS} has a run time of approximately 4.5 minutes under the same conditions. This allowed us to use a model grid of a much greater size. As this approach removes the rolling aspect of {r}RMS, the improved algorithm was renamed cross RMS (xRMS).

In addition to the increase in speed, an estimate of the error on the recovered thermodynamic properties may be found. To do this, we saved the 20 next-best models after the best model. The choice of 20 was decided in order to fit within the computational limits. With enough memory, xRMS can be asked to hold any arbitrary number of `next-best' models. These models were subject to the RMS threshold. If they were above the RMS threshold, they were discarded. The surviving models were then used to make maps of the thermodynamic properties of those models. These maps were then subtracted from the `best' map, and the map that produced the largest difference was chosen as the error. This was done on a pixel-by-pixel basis, and the model chosen to represent the error in one pixel is not necessarily the same as its neighbouring pixel. These models can be plotted in a similar manner to the diagnostic maps. {Pixels that do no have any `next-best' matches have an error of $\pm$0 assigned. To avoid confusion, we assigned any pixels where no `best-match' was found a value of NaN, which is plotted as a bad pixel (white).}

\section{Results}

\begin{figure}
    \centering
    \includegraphics[width=.49\linewidth]{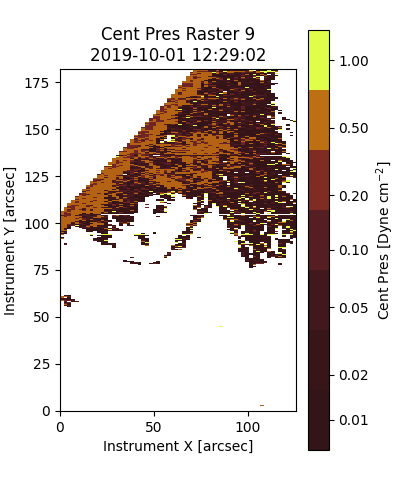}
    \includegraphics[width=.49\linewidth]{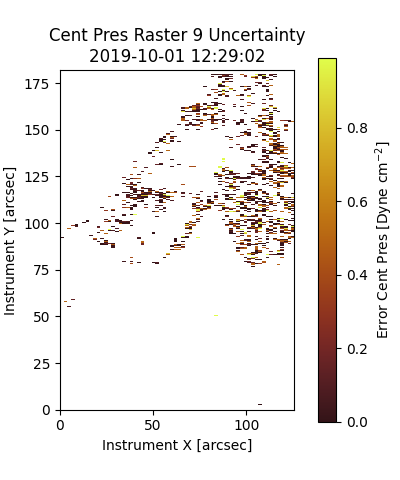}
    \includegraphics[width=.49\linewidth]{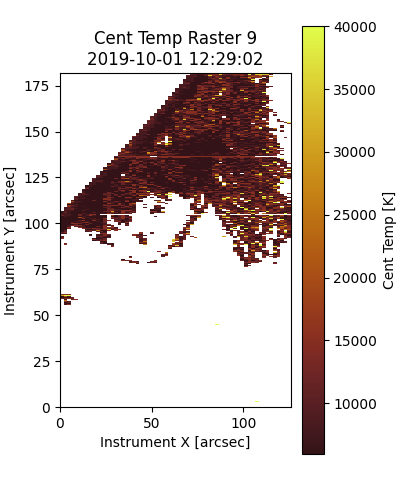}
    \includegraphics[width=.49\linewidth]{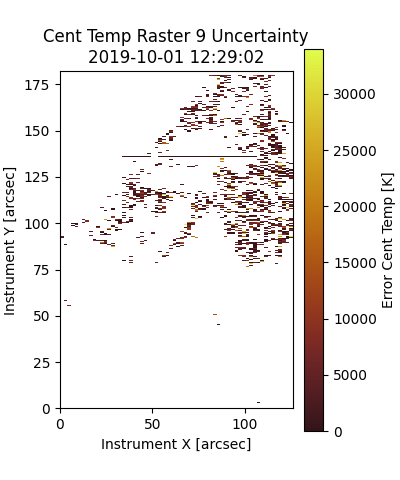}
    \includegraphics[width=.49\linewidth]{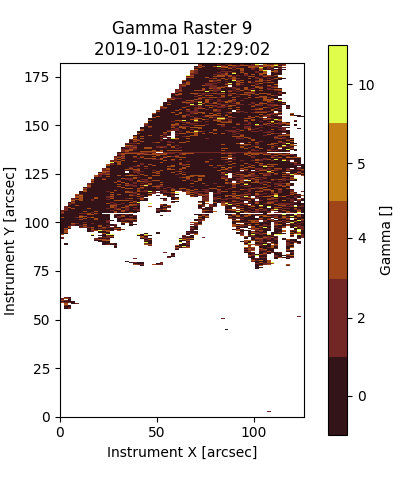}
    \includegraphics[width=.49\linewidth]{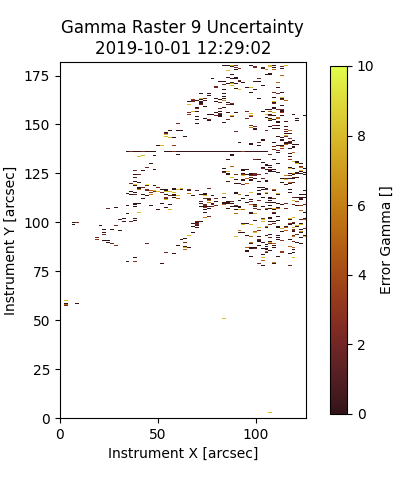}
    \caption{Select diagnostics of the {ninth} raster of the prominence with their associated fractional errors. The uncertainty plots only show uncertainties where good matches are found.}
    \label{fig:20191001_1}
\end{figure}

\begin{figure}
    \centering
    \includegraphics[width=.49\linewidth]{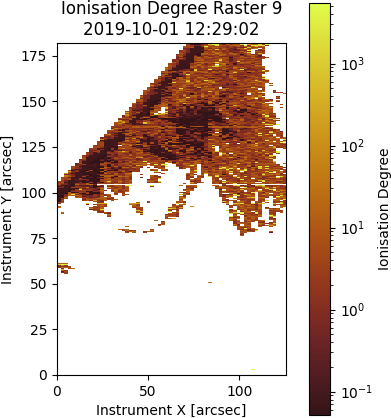}
    \includegraphics[width=.48\linewidth]{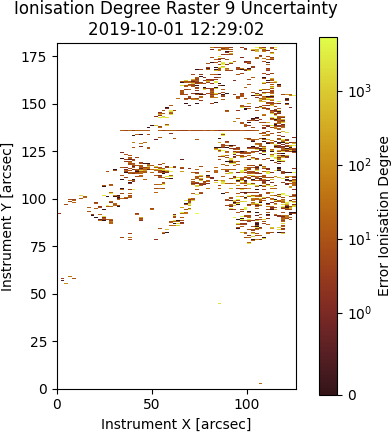}
    \includegraphics[width=.49\linewidth]{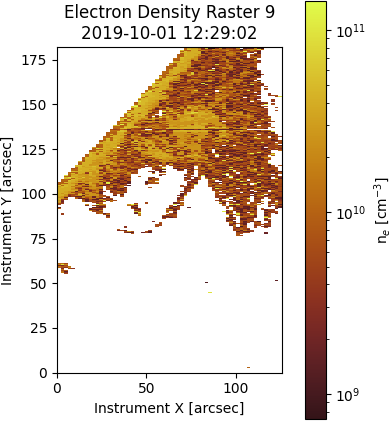}
    \includegraphics[width=.48\linewidth]{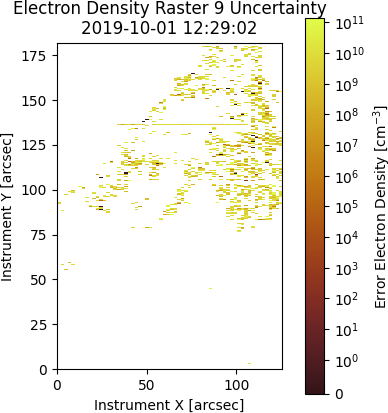}
    \caption{Electron density and ionisation degree of the {ninth} raster of the prominence with their associated errors. The uncertainty plots only show uncertainties where good matches are found.}
    \label{fig:20191001_2}
\end{figure}

\begin{figure}
    \centering
    \includegraphics[width=.54\linewidth]{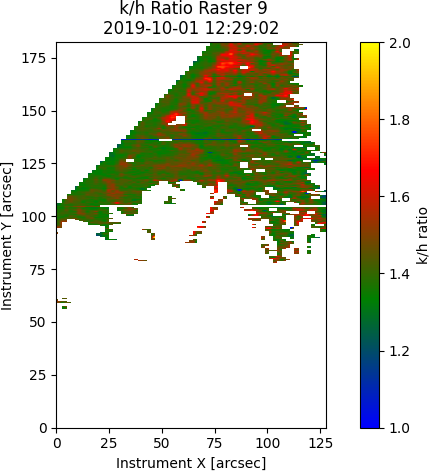}
    \includegraphics[width=.43\linewidth]{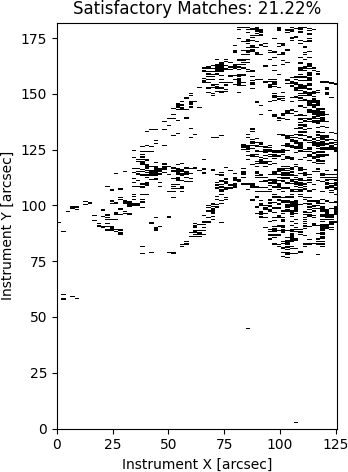}
    \caption{\ion{Mg}{ii} k/h ratio and satisfactory matches. No clear correlation is present.}
    \label{fig:goodbadkh}
\end{figure}

{Figures \ref{fig:20191001_1} and \ref{fig:20191001_2} show a selection of the diagnostics recovered by xRMS for raster 9. {Satisfactory and unsatisfactory matches are both presented in these diagnostic maps for readability, and the maps should be read in tandem with the satisfactory matches plot (see right of Fig. \ref{fig:goodbadkh}).} {The errors here are given as simple plus-minus errors; however, it is important to note that single ion prominence inversions produce multi-peaked posterior likelihoods \citep{peat_inprep2}, so not every number in the stated range is an acceptable value.} 

Raster 9 was chosen as our example, as it {simply} has the greatest number of satisfactory matches, 21.22\%. The majority of our satisfactory matches are towards the edges of the prominence, where {{less fine structure}} is expected to be observed.} As can be seen, the {central} pressure of the prominence is quite low, remaining below 0.2$~$dyn$~$cm$^{-2}$. However, these pixels are unsatisfactory matches (see Fig. \ref{fig:goodbadkh}, right). The uncertainty map shows errors of {approximately $\pm$0.2 to $\pm$0.3$~$dyn$~$cm$^{-2}$}. This could be due to solutions where {the effects of pressure} can instead {be produced by} some other parameter.

The observed {central} temperatures are {rather} high {compared to typical prominences. Here, we have values}  ranging from {approximately} 10~kK to {20~kK}. This is consistent with {the mean temperatures recovered in} our previous study using rRMS \citep{peat_solar_2021}. However, the lower bound found in our previous study was 6~kK. {We found errors of up to approximately {5~kK to 15~kK} in the areas with good matches. This means the temperature can be up to twice as high or twice as low as that recovered by the best match}. {This disparity in solutions is likely due to the same reason given for the differences in central pressure.}  This may demonstrate the degeneracies faced when using a single ion and/or species from 1D prominence modelling to infer the thermodynamic properties \citep{jejcic_2022, peat_inprep2}. 

{In total, we recovered 14.1\% satisfactory matches{, 50.3\% of which were PCTR models}. This is in stark contrast to the 49\% satisfactory matches recovered in \cite{peat_solar_2021}. The main difference between these two observations is the orientation of the prominence. Here, the line of sight (LOS) is almost parallel with the spine, and in the aforementioned study, the LOS was almost perpendicular to the spine. In the former orientation, we would expect to intercept more {{fine structure}}, which can lead to complicated line profiles that 1D prominence modelling cannot effectively reproduce \citep{gunar_large_2022, peat_mg_2023}.}
{The gamma plots allowed us to investigate the structure of the PCTR. Where we did find good matches, {{the gamma factor appears to flip}} between two, four, and five. The gamma error plot not only shows the uncertainty where a PCTR model is selected as the best fit but also further backs up {{the flipping nature of the gamma factor}}, as the uncertainties are between one and three. This shows that there is a relatively gentle PCTR gradient.}

As in the previous study, we recovered a large ionisation degree. We used the definition of ionisation degree from \cite{tandberg-hanssen_nature_1995},
\begin{equation}
    \text{I.D.}=\frac{n\ion{H}{ii}}{n\ion{H}{i}},
\end{equation}
as utilising the commonly used approximations of $n$\ion{H}{ii}$\approx n_e$ and $n$\ion{H}{i}$\approx n\text{H}_\text{0}$ (where $n\text{H}_\text{0}$ is {{the number density of ground state hydrogen}}) can overestimate $n$\ion{H}{ii} by up to 20\% \citep{peat_solar_2021}. Our recovered ionisation degree appears quite reasonable, with the largest values appearing at the edges of the prominence material, where one would expect to encounter more of the PCTR. Here, the errors mainly range from $\pm$1 to $\pm$100. The electron density map displays a similar distribution, with more electrons closer to regions that intersect the PCTR. The majority of the structure appears to exhibit an electron density of {$10^{10}$~cm$^{-3}$}. The {majority of the pixels in the} uncertainty {plot appear to be around} {$\pm$10$^9$ to $\pm$10$^{10}~$cm$^{-3}$.}

Unlike in \citet{peat_solar_2021}, we did not find a weak objective correlation between the areas where we found satisfactory matches and the regions with a high k/h ratio (see Fig. \ref{fig:goodbadkh}). {The generally low k/h ratio could also be why the algorithm did not recover many satisfactory matches}. In the optically thin regime, the k/h ratio becomes that of the oscillator strengths. The oscillator strengths of \ion{Mg}{ii}~h\&k are 0.300 and 0.601, respectively \citep{theodosiou_accurate_1999}, giving a ratio of approximately two. {Due to the large amount of potential fine structure along the LOS, it is likely that the best matches are found where the optical thickness is the highest, as this is where we only observed the `front-most' structure.} However, due to the scattering of incident radiation, the k/h ratio can exceed two in solar prominences \citep{heinzel_notitle_2022}. Therefore, this may not be a reliable measurement of the optical thickness.

\section{Conclusions}
    The prominence of 1 October 2019 was very dynamic, with one large flow observed towards the solar limb. From the line statistics recovered via the quantile method, we found the prominence to contain a mixture of asymmetries between 0.8 and -0.8. It was fairly dynamic, with recovered LOS velocities of up to 70~km~s$^{-1}$.  Through the use of the new xRMS algorithm, we found {central} temperatures of 10~kK to {20~kK} and {central} pressures of 0.2~dyn~cm$^{-2}$. The ionisation degree was found to range from around 1 to {100}, with an electron density of mainly {$10^{10}$~cm$^{-3}$}. The range of the associated uncertainties recovered via this new method are quite large, {an order of magnitude so}. This is likely an example of the degeneracies faced when using single ion and/or species 1D prominence inversions \citep{jejcic_2022, peat_inprep2}.  In future work, the algorithm should be expanded to have the ability to match several co-aligned observations in different ion and/or species (e.g. \ion{Mg}{ii}~h\&k and H$\alpha$) to attempt to constrain these degenerate solutions. {Additionally, prominences showing higher k/h ratios and/or oriented perpendicular to the LOS may also be better candidates for this method.}
   
\begin{acknowledgements}
    AWP acknowledges financial support from STFC via grant ST/S505390/1. NL acknowledges support from STFC grant ST/T000422/1. 
    We would also like to thank Dr Graham S. Kerr for supplying the IRIS spatial PSFs and Dr Christopher M. J. Osborne for writing the Python implementation of WENO4.
    IRIS is a NASA small explorer mission developed and operated by LMSAL with mission operations executed at NASA Ames Research Center and major contributions to downlink communications funded by ESA and the Norwegian Space Centre. AIA data courtesy of NASA/SDO and the AIA, EVE, and HMI science teams. 
    Hinode is a Japanese mission developed and launched by ISAS/JAXA, collaborating with NAOJ as a domestic partner, NASA and STFC (UK) as international partners. Scientific operation of the Hinode mission is conducted by the Hinode science team organized at ISAS/JAXA. This team mainly consists of scientists from institutes in the partner countries. Support for the post-launch operation is provided by JAXA and NAOJ (Japan), STFC (U.K.), NASA (U.S.A.), ESA, and NSC (Norway). 
    This research used version 3.7.1 of Matplotlib \citep{matplotlib}, version 1.24.3 of NumPy \citep{harris_array_2020}, version 1.10.1 of SciPy \citep{scipy}, version 4.1.5 of the SunPy open source software package \citep{barnes_sunpy_2020}, version 0.7.3 \citep{aiapysoft} of the aiapy open source software package \citep{aiapy}, and version 5.2.2 of Astropy (http://www.astropy.org) a community-developed core Python package for Astronomy \citep{the_astropy_collaboration_astropy_2013, the_astropy_collaboration_astropy_2018}.
\end{acknowledgements}

\bibliography{bibtex}
\bibliographystyle{aa}

\appendix
\begin{figure*}
    \section{\ion{Mg}{ii}~k statistic maps}
    \centering
    \resizebox{0.8\hsize}{!}
    {\includegraphics{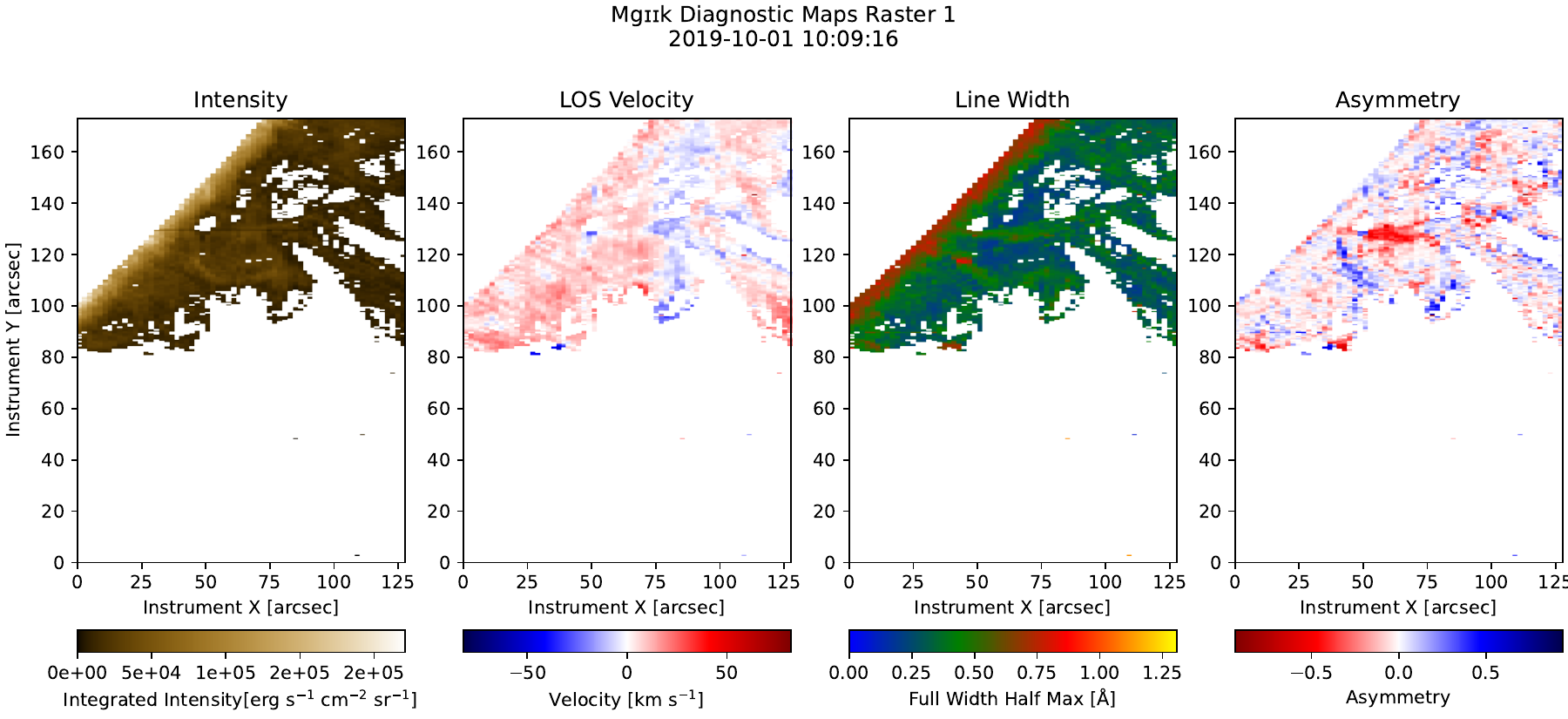}}
    \resizebox{0.8\hsize}{!}
    {\includegraphics{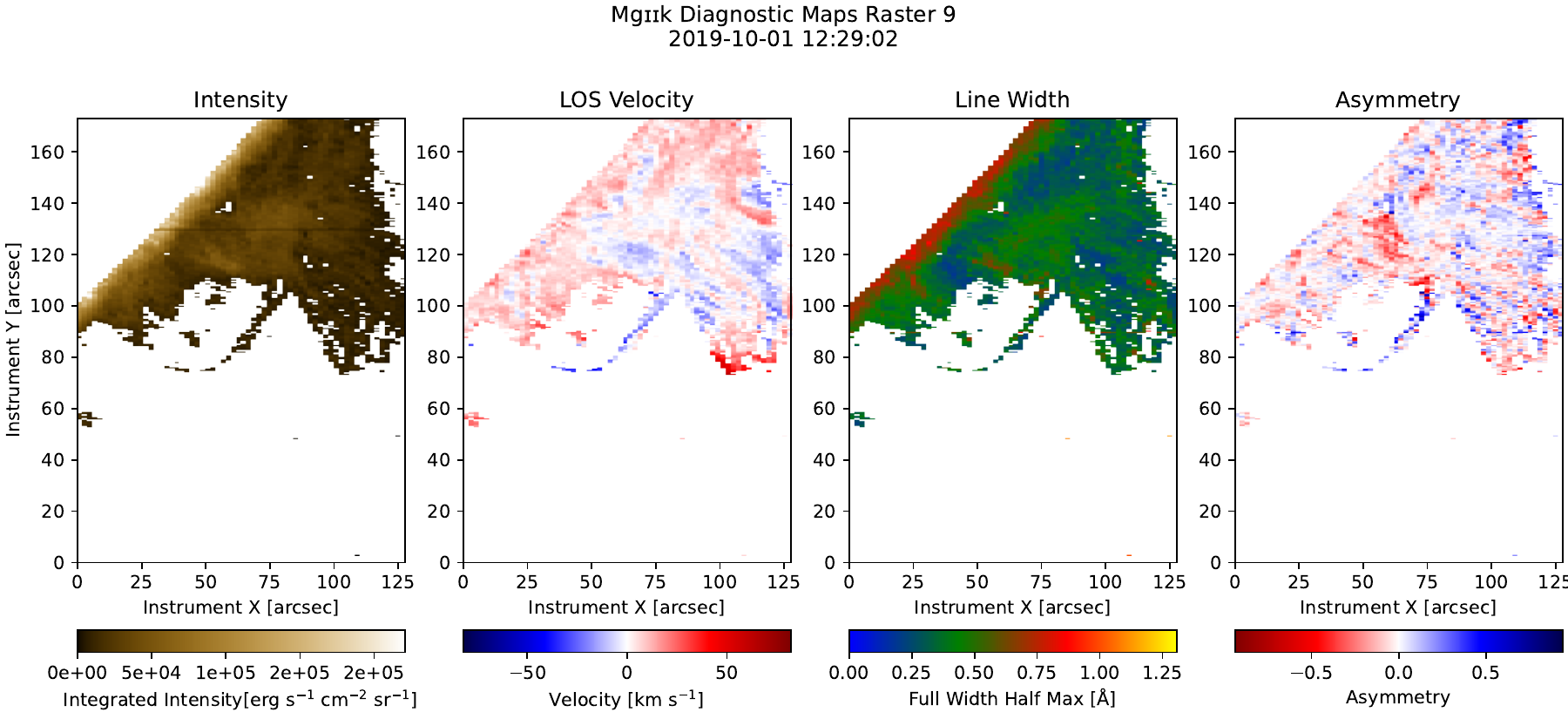}}
    \resizebox{0.8\hsize}{!}
    {\includegraphics{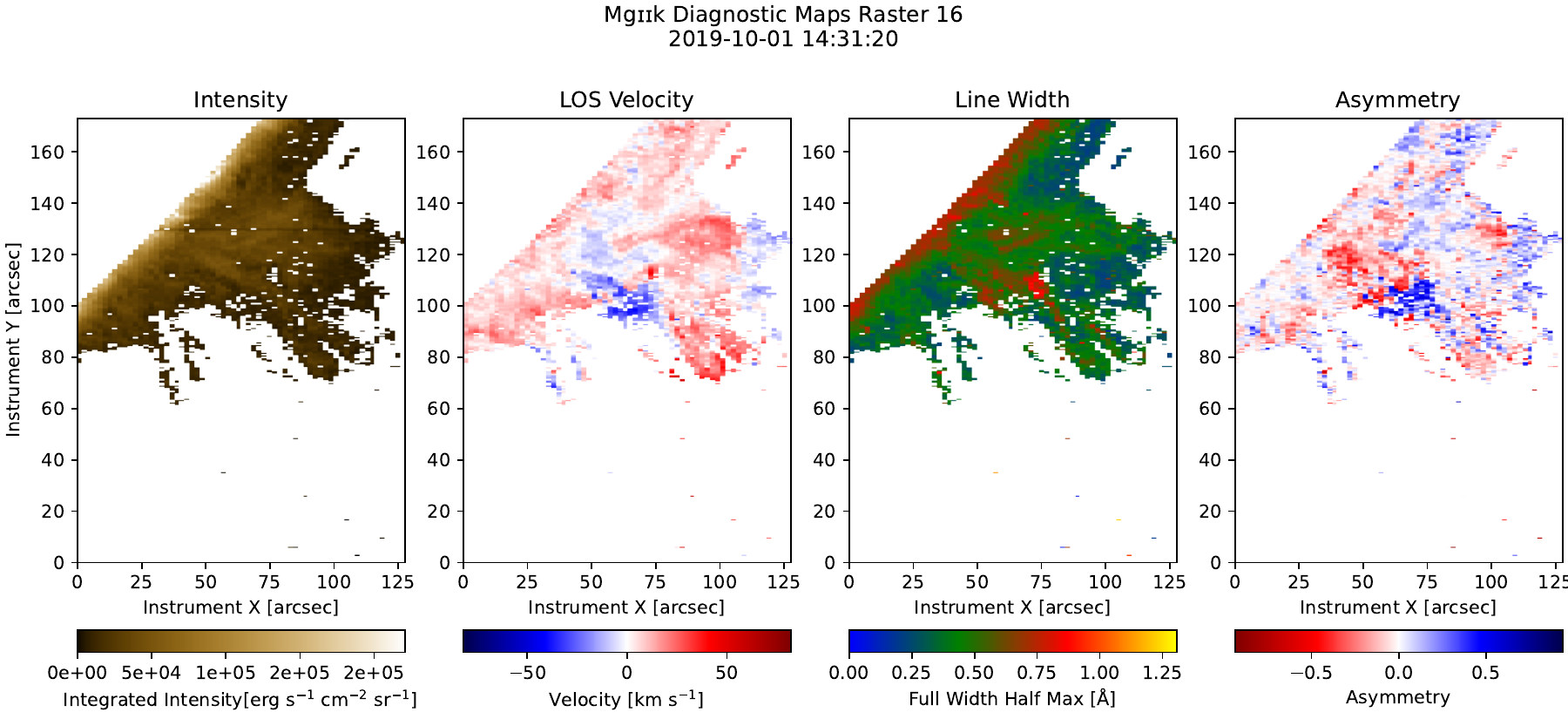}}
    \caption{Same as Fig. \ref{maps} but for \ion{Mg}{ii}~k.}
    \label{kmaps}
\end{figure*}

\end{document}